\documentclass[singlecolumn,superscriptaddress,english,pre,showpacs,longbibliography]{revtex4-2}

\usepackage[colorlinks=true,urlcolor=blue,citecolor=blue,linkcolor=blue]{hyperref} 
\usepackage{graphicx}
\usepackage{svg}
\usepackage{amssymb}
\usepackage{dcolumn}
\usepackage{bm}
\usepackage{color}
\usepackage{physics}
\usepackage{float}

\usepackage{algorithm}
\usepackage{algorithmic}

\begin{document}


\title{Integrating Neural Networks and Tensor Networks for Computing Free Energy}

\author{Hanyan Cao}
\affiliation{
CAS Key Laboratory for Theoretical Physics, Institute of Theoretical Physics, Chinese Academy of Sciences, Beijing 100190, China
}
\affiliation{
 School of Physical Sciences, University of Chinese Academy of Sciences, Beijing 100049, China
}
\author{Yijia Wang}
\affiliation{
CAS Key Laboratory for Theoretical Physics, Institute of Theoretical Physics, Chinese Academy of Sciences, Beijing 100190, China
}
\affiliation{
 School of Physical Sciences, University of Chinese Academy of Sciences, Beijing 100049, China
}

\author{Feng Pan}
\affiliation{
CAS Key Laboratory for Theoretical Physics, Institute of Theoretical Physics, Chinese Academy of Sciences, Beijing 100190, China
}
\affiliation{Centre for Quantum Technologies, National University of Singapore, 117543, Singapore}
\affiliation{Science, Mathematics and Technology Cluster, Singapore University
of Technology and Design, 8 Somapah Road, 487372 Singapore}

\author{Pan Zhang}
\email{panzhang@itp.ac.cn}
\affiliation{
 CAS Key Laboratory for Theoretical Physics, Institute of Theoretical Physics, Chinese Academy of Sciences, Beijing 100190, China
}
\affiliation{School of Fundamental Physics and Mathematical Sciences, Hangzhou Institute for Advanced Study, UCAS, Hangzhou 310024, China}
\affiliation{International Centre for Theoretical Physics Asia-Pacific, Beijing/Hangzhou, China}


\begin{abstract}
Computing free energy is a fundamental problem in statistical physics. Recently, two distinct methods have been developed and have demonstrated remarkable success: the tensor-network-based contraction method and the neural-network-based variational method. Tensor networks are accurate, but their application is often limited to low-dimensional systems due to the high computational complexity in high-dimensional systems. The neural network method applies to systems with general topology. However, as a variational method, it is not as accurate as tensor networks. In this work, we propose an integrated approach, tensor-network-based variational autoregressive networks (TNVAN), that leverages the strengths of both tensor networks and neural networks: combining the variational autoregressive neural network’s ability to compute an upper bound on free energy and perform unbiased sampling from the variational distribution with the tensor network’s power to accurately compute the partition function for small sub-systems, resulting in a robust method for precisely estimating free energy.
To evaluate the proposed approach, we conducted numerical experiments on spin glass systems with various topologies, including two-dimensional lattices, fully connected graphs, and random graphs. Our numerical results demonstrate the superior accuracy of our method compared to existing approaches. In particular, it effectively handles systems with long-range interactions and leverages GPU efficiency without requiring singular value decomposition, indicating great potential in tackling statistical mechanics problems and simulating high-dimensional complex systems through both tensor networks and neural networks.

\end{abstract}

\maketitle


\section{\label{sec:level1} Introduction}
The computation of free energy is a cornerstone of statistical physics, serving as a critical tool for understanding equilibrium properties, phase transitions, and macroscopic behaviors in complex systems~\cite{landau2013statistical}. Accurate free energy estimation enables the prediction of thermodynamic quantities and the exploration of emergent phenomena in diverse systems, from ordered lattices to disordered spin glasses~\cite{sethna2021statistical}. However, for finite-size systems, computing the exact free energy is a $\#$P-hard problem due to the exponential summation over variable configurations. To address this challenge, a series of algorithms based on the variational method have been proposed such as Mean Field approximation~\cite{jordan1999introduction}, Bethe ansatz~\cite{bethe1935statistical} and Kikuchi loop expansions~\cite{PhysRev.81.988}. These approaches adopt different variational ansatz for the joint distribution $q_{\theta}$ with variational parameter $\theta$ and optimize the Kullback–Leibler (KL) divergence~\cite{mackay2003information} between $q_{\theta}$ and the Boltzmann distribution $p$. Utilizing the non-negativity of KL-divergence~\cite{mackay2003information}, these methods can obtain an upper bound on the free energy. Nevertheless, on systems with correlations of different scales and random sparsity, the performance of the above methods falls short of expectations. 

Two prominent computational approaches have emerged in recent years: tensor network (TN) methods~\cite{levin2007tensor} and neural network (NN)-based variational methods~\cite{PhysRevLett.122.080602}. Tensor networks excel in low-dimensional systems, offering high precision by exploiting the locality and sparsity of interactions to estimate the partition function efficiently. Yet, their application is hindered by prohibitive computational complexity in high dimensions or systems with long-range correlations. Conversely, variational neural networks, based on autoregressive models, provide a flexible framework for handling systems of arbitrary topology. This method adopts an autoregressive variational ansatz for the joint distribution, enabling parallel sampling and gradient-based optimization. However, their accuracy is inherently limited by the system scale and the difficulty of the optimization of neural networks.

To bridge these gaps, we propose Tensor-Network-based Variational Autoregressive Networks (TNVAN), a hybrid framework that synergizes the complementary strengths of tensor networks and neural networks. TNVAN  decomposes the system into two subsets. The smaller and denser subset is named the width set, which contributes critically to the growth of the contraction complexity of tensor networks measured by the contraction width. Building upon the concept of the Feedback Vertex Set (FVS)~\cite{PhysRevE.103.012103}, the removal of variables in the width set generalizes the remaining graph from a forest that can be explicitly computed to a graph with more complex structures that can be computed within acceptable complexity. After decomposing the system and fixing the variables in the width set, the method projects the whole system onto the width set by contracting the tensor network of the remaining subsystem, which has lower complexity. For the equivalent system containing only the width set, TNVAN utilizes the variational autoregressive network (VAN) ~\cite{PhysRevLett.122.080602} to efficiently estimate the free energy.

Our new approach leverages the variational autoregressive network to compute an upper bound on the free energy and perform unbiased sampling, while employing tensor networks to precisely evaluate the contribution of the reduced subsystems to the partition function. This integration circumvents the high-dimensional limitations of pure TN methods and mitigates the accuracy trade-offs of purely variational approaches.  Numerical experiments on spin glass systems across 2D lattices, random graphs, and fully connected networks demonstrate TNVAN's superior accuracy and scalability, particularly in regimes with long-range interactions. This unified approach paves the way for tackling high-dimensional statistical mechanics problems and simulating complex systems where traditional methods fall short.

The rest of the article is organized as follow. In Sec.~\ref{sec:2}, we review the methodology of tensor networks for computing the free energy, i.e., the partition function, and describe how to project the whole system onto its subsystem composed of the width set.  Sec.~\ref{sec:3} explains how the variational autoregressive network is used to estimate the free energy by sampling from the variational distribution and optimizing the variational free energy. In Sec.~\ref{sec:4}, we show numerical results on statistical models with different topological structures and parameter settings. These numerical results demonstrate TNVAN's superior accuracy in computing free energy for various spin systems compared to other variational methods. Finally, Sec.~\ref{sec:5} concludes the paper, summarizing the advantages of TNVAN and outlining potential future work to further enhance its applicability and accuracy.

\begin{figure*}[t]
\centering
    \includegraphics[width=0.9\linewidth]{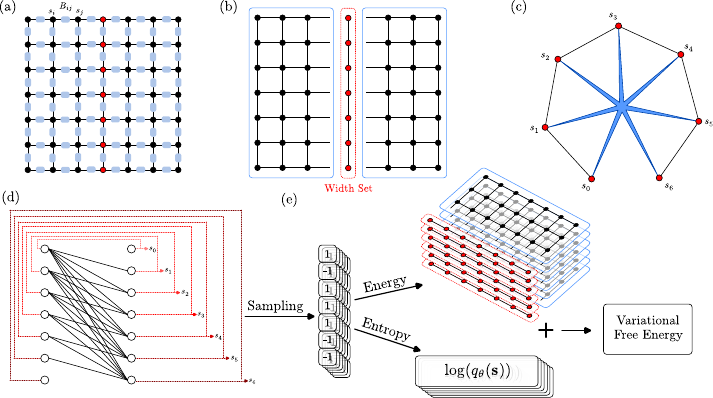}
    \caption{TNVAN method. (a) The 2D Ising model on a $7\times7$ lattice. The tensor network consists of hyper-indices (representing spins) and tensors (representing Boltzmann matrices). The partition function of this spin model is equal to the result of the tensor network contraction with the same topology. (b) The width set (red nodes) and the remaining variables (black nodes). When the configurations of variables in the width set are fixed, the space complexity of the remaining tensor network is reduced to half of that of the original tensor network. (c) The equivalent spin model with $7$ spins. The blue heptagram represents the extra energy from the reduced TN. (d) The structure of the neural network which is named as Masked Autoencoder for Distribution Estimation (MADE)~\cite{germain2015made}. The dotted line represents the data flow during the sampling process. The color from light to dark indicates the order of sampling. (e) The process for computing the variational free energy. Batch samples are drawn, and the energy and entropy are calculated respectively to estimate the free energy.}
    \label{fig:1}
\end{figure*} 
    
\section{Computing free energy by tensor network contraction}
\label{sec:2}
Consider a statistical physics system such as the celebrated Ising model with $n$ spins and an energy function $E(\boldsymbol{s})=-\sum_{\langle ij\rangle}J_{ij}s_is_j$, $\boldsymbol{s}\in\{+1, -1\}^n$, the joint probability of $\boldsymbol{s}$ follows the Boltzmann distribution:
\begin{equation}
    p(\boldsymbol{s}) = \frac{1}{Z}\mathrm{e}^{-\beta E(\boldsymbol{s})}\;,
\end{equation}
where $\beta$ is the inverse temperature and $Z=\sum_{\boldsymbol{s}} \mathrm{e}^{-\beta E(\boldsymbol{s})}$ is the partition function. The free energy is defined as $F = - \frac{1}{\beta}\ln{Z}$. From the definition of the partition function, its calculation can be viewed as the contraction of a tensor network~\cite{Pan2019}:
\begin{equation} Z=\sum_{\boldsymbol{s}}\mathrm{e}^{-\beta E(\boldsymbol{s})}=\sum_{\boldsymbol{s}}\prod_{\langle ij\rangle} B_{ij}(s_i, s_j)\;,
\end{equation}
where 
$B_{ij}(s_i, s_j)$ is the Boltzmann matrix with $B_{ij}(+,+) = B_{ij}(-,-) = \mathrm{e}^{\beta J_{ij}}$ and $B_{ij}(+,-) = B_{ij}(-,+) = \mathrm{e}^{-\beta J_{ij}}$. The tensor network $\mathbf{TN}(\{B\}, \{s\})$ for calculating the partition function of this system can be represented by the same diagram, where the spins (vertices) represent the indices and the edges represent the Boltzmann matrices as illustrated in Fig.~\ref{fig:1} (a). However, computing the partition function or contracting the tensor network $\mathbf{TN}(\{B\}, \{s\})$ is $\#$P-hard in general due to the exponential configuration space. An effective approach to solving this problem is to divide the whole system into a tree(or forest) and another subsystem named Feedback Vertex Set~\cite{festa2012feedback, PhysRevE.103.012103}, with a smaller number of variables compared to the original system. When the configurations of variables in FVS have been fixed, the remaining tree is easy to reduce from the leaves to the roots. Based on this idea we also divide the variables of the whole system into two parts $\{s\} =\mathcal{W} \cup \mathcal{R}$
as shown in Fig.~\ref{fig:1} (b), and we name $\mathcal{W}$ as the width set. The width set needs to satisfy that the contraction width $w(\mathbf{TN(\mathcal{R})})$ of the remaining tensor network is smaller than an upper bound $w_u$ after fixing all variables in $\mathcal{W}$. The contraction width of a tensor network is the minimum space complexity across all possible contraction orders and the space complexity is the maximum dimension of intermediate tensors during the contraction process~\cite{Gray2021hyperoptimized}.

At this time the partition function can be rewritten as following form:
\begin{equation}
    Z = \sum_{\boldsymbol{s}\in\mathcal{W}}\sum_{\boldsymbol{r}\in\mathcal{R}}\mathrm{e}^{-\beta E(s, r)} = \sum_{\boldsymbol{s}\in\mathcal{W}}\mathrm{e}^{-\beta \tilde{E}(\boldsymbol{s})}\;,
\end{equation}
where $\boldsymbol{s}\in\{+1, -1\}^{|\mathcal{W}|}$ and $\boldsymbol{r}\in\{+1, -1\}^{|\mathcal{R}|}$ are configurations of variables in $\mathcal{W}$ and $\mathcal{R}$ separately. The effective energy has two terms:
\begin{equation}
   \tilde{E}(\boldsymbol{s}) = E_1(\boldsymbol{s}) - \frac{1}{\beta} \ln{\sum_{\boldsymbol{r}\in \mathcal{R}}} \mathrm{e}^{-\beta E_2(\boldsymbol{s}, r)}\;,
\end{equation}
where $E_1$ represents the energy of the internal interactions among $\boldsymbol{s}$ and $E_2$ represents the energy of the interactions involving the variables in $\mathcal{R}$ which needs to be summed over.

As shown in Fig.\ref{fig:1} (b), when the configurations of the variables in $ \mathcal{W}$, which is the middle column of the lattice, are fixed, the remaining tensor network can be contracted with complexity within the predefined acceptable range. By tracing out the spins in $\mathcal{R}$ and treating their effects as an extra interaction among the width set variables, the system is reduced to a much smaller effective system that contains only the spins in $ \mathcal{W}$ as Fig.~\ref{fig:1} (c). By enumerating all possible configurations of variables in $\mathcal{W}$, we can get the exact contraction results. This method for tensor network contraction is known as slicing, variable projection~\cite{chen2018classical}, or bond cutting~\cite{villalonga2019flexible}. It has been widely used in quantum circuit simulation~\cite{Pan2022, Pan2021}. 

However, finding the width set is an NP-hard computational task~\cite{xu2023nphardnesstensornetworkcontraction} requires finding a contraction order $\pi$ with the minimum space complexity by definition. A series of heuristic algorithms have been developed to address this challenge, such as Greedy~\cite{kask2011pushing}, Simulate Annealing (SA)~\cite{kalachev2022} and other optimization methods based on contraction trees~\cite{Gray2021hyperoptimized}. After determining a good contraction order $\pi$  for the original tensor network, we employ an algorithm called dynamic slicing~\cite{chen2018classical, villalonga2019flexible, Gray2021hyperoptimized} to identify the width set. This algorithm constructs a contraction tree $\mathcal{T}$ based on the order $\pi$. It then iteratively removes the index (spin) with the maximum width and updates $\mathcal{T}$ until $w(T)<w_u$. These removed indices are as width set variables. Once the width condition is satisfied, we randomly remove indices from $\mathcal{W}$ and add them back to $\mathcal{T}$. In this work, we utilize the Cotengra~\cite{Gray2021hyperoptimized} package as well as another simulated annealing algorithm implemented by ourselves to determine
the width set and the corresponding order for the remaining tensor network.

\section{Estimating the free energy using VAN}
\label{sec:3}

As described in the previous section, we have used tensor networks to project a complete system onto an equivalent system with a smaller number of variables and denser interactions. At this time, exhausting the configurations of the variables in $\mathcal{W}$ yields an exact free energy, but this approach tends to fail when the number of variables in $\mathcal{W}$ increases. Here we introduce a method based on the variational method, which adopts a variational distribution $q$ and estimates the free energy by minimizing the reverse KL-divergence $\mathrm{D}_{\mathrm{kl}}(q|p)$ between $q_{\theta}(\boldsymbol{s})$ and the Boltzmann distribution $p(\boldsymbol{s}) = \frac{\mathrm{e}^{-\beta \tilde{E}(\boldsymbol{s})}}{Z}$, which is nothing but the difference between the variational free energy and the true free energy.
\begin{equation}
\begin{split}
    \mathrm{D}_{\mathrm{kl}}(q_\theta|p) &= \sum_s q_{\theta}(\boldsymbol{s})\left\{\ln{q_\theta(\boldsymbol{s})}-\ln{p}(\boldsymbol{s})\right\}\\
    &= \sum_s q_{\theta}(\boldsymbol{s})\left\{\ln{q_\theta(\boldsymbol{s})}+\beta\tilde{E}(\boldsymbol{s})\right\} -\ln{Z}\\
    &=\beta(F_q-F)\;.
\end{split} 
\end{equation}
Due to the non-negativity of $\mathrm{D}_{\mathrm{kl}}(q|p)$, it can be shown that the variational free energy is an upper bound of the true free energy. And optimizing it with gradient descent is actually optimizing the variational free energy $F_q$ (this is because the true free energy does not contain the parameter $\theta$).

The most commonly used variational method is the Mean-Field approximation, which assumes that the joint distribution is the product distribution $q(\boldsymbol{s}) = \prod_i q(s_i)$. In this work, we adopt a more expressive variational ansatz utilizing autoregressive neural networks~\cite{germain2015made, pmlr-v70-kalchbrenner17a, Vaswani2017} to estimate the free energy of the width set system. These neural networks are widely used in machine learning to process sequential data, such as natural language, audio, images, etc. In statistical physics, these neural networks are used to parameterize the Boltzmann distribution by representing the joint distribution of variables as the product of conditional distributions:
\begin{equation}
    q_{\theta}(\boldsymbol{s}) = \prod_i q(s_i|s_{j<i})\;,
\end{equation}
where $s_{j<i}$ represents the variables with indices smaller than $i$ in a predetermined order. One can see that the state of $s_i$ only depends on the variables before it. The network structure is shown in Fig.~\ref{fig:1} (d). We use configurations of variables as input and obtain the probability of spin up $\hat{s}(s_i=+1|s_{j<i})$ of each variable as output:
\begin{equation}
    \hat{s}_i = \sigma(\sum_{j<i} W_{ij}\cdot s_j +b_i)\;,
\end{equation}
where $\sigma$ is the \textbf{Sigmoid} function and $W_{ij}$ and $b_i$ is the weights and bias of the neural network. Then the conditional distribution of each variable is:
\begin{equation}
    q(s_i|s_{j<i}) = \hat{s_i}\delta_{s_i, +1} + (1-\hat{s_i})\delta_{s_i, -1}\;,
\label{eq:8}
\end{equation}
which is nothing but the Bernoulli distribution. We
can directly draw samples from the variational distribution~\cite{PhysRevLett.122.080602}. We can draw samples for the the first spin $s_1$ from $q(s_1)$ with probabilities $q(s_1=+1) = \hat{s_1}$ and $q(s_1=-1) = 1-\hat{s_1}$ . Then, we can draw samples of $s_2$ based on $q(s_2|s_1)$ by inputting the configuration of $s_1$. Sequentially, inputting the configurations of $s_1$ and $s_2$ into the neural network, we can obtain the configuration of $s_3$ based on the conditional distribution $q(s_3|s_2, s_1)$, and so on until the configuration of the last spin $s_7$ is known. Using these samples we directly compute the variational free energy as illustrated in Fig.~\ref{fig:1} (e)
\begin{equation}
\label{eq:9}
    F_q = \mathbb{E}_{\boldsymbol{s}\sim q_{\theta}(\boldsymbol{s})} \left[ \tilde{E}(\boldsymbol{s}) + \frac{1}{\beta}\ln{q_{\theta}(\boldsymbol{s})}\right]\;.
\end{equation}
This variational free energy consists of the effective energy and the entropy for variables in the width set. For the effective energy, we using the batch-contraction to contract multiple tensor networks and average over a batch of samples. For the entropy, we use the output of the neural network to compute the log probabilities. To minimize the variational free energy, we compute the gradient of the variational free energy and use gradient descent to approximate its 
upper bound
\begin{equation}
\label{eq:10}
\begin{split}
    &\nabla_{\theta} F_{q} = \\
    &\mathbb{E}_{\boldsymbol{s}\sim q_{\theta}(\boldsymbol{s})} \left\{\left[\tilde{E}(\boldsymbol{s}) + \frac{1}{\beta}\ln{q_{\theta}(\boldsymbol{s})} \right]\nabla_{\theta}\ln{q_{\theta}(\boldsymbol{s})}\right\}\;,
\end{split}
\end{equation}
which is well known in reinforcement learning~\cite{sutton2018reinforcement,avramidis1996integrated}. This gradient is utilized to iteratively update the parameters of the neural network through gradient descent optimization until the variational free energy converges to a stable minimum. Once convergence is achieved, we can efficiently draw unbiased samples from the trained neural network, which approximates the Boltzmann distribution of the system. These samples enable the calculation of various physical quantities, such as magnetization, spin-spin correlations, and susceptibility, providing a comprehensive understanding of the system's thermodynamic properties~\cite{PhysRevE.103.012103}. We name this innovative method tensor-network-based variational autoregressive networks, reflecting its unique integration of tensor networks and neural networks. And the pseudo code of free energy estimation using TNVAN is shown in Algorithm~\ref{alg:1}. TNVAN represents a significant advancement over traditional approaches by combining the high precision of tensor networks in handling local interactions with the flexibility and scalability of neural networks in modeling complex, high-dimensional systems. This hybrid framework not only improves the accuracy of free energy estimation but also extends the applicability of variational methods to a broader range of physical systems.

In fact, TNVAN can be viewed as a generalization of the Feedback-Vertex-Set variational autoregressive network (FVS-VAN)~\cite{PhysRevE.103.012103}, which relies on the concept of the Feedback Vertex Set to decompose systems into FVS and tree-like structures. TNVAN extends this idea by introducing the more flexible concept of the width set. The width set is not limited to the fact that the remaining system has a tree-like structure, instead, it allows for the remaining graph to have more complex topologies, provided that the contraction complexity remains manageable. The key distinction lies in how the effective energy for the width set is computed. TNVAN employs tensor network contractions to accurately capture the contributions of the remaining subsystem, whereas FVS-VAN relies on analytical reductions for tree structures. This generalization enables TNVAN to handle a wider variety of systems, including those with long-range interactions and non-tree-like correlations, making it a more versatile and powerful tool for statistical mechanics simulations.

\begin{algorithm}[H]
\caption{TNVAN for free energy estimation.}
\begin{algorithmic}
\label{alg:1}
\REQUIRE The tensor network of Ising model $\mathbf{TN}(\{B\}, \{s\})$; the upper bound of contraction width $w_u$; 
\ENSURE The variational free energy $F_q$
\STATE 1. Using the existing method to obtain $\mathcal{W}$ and the contraction order $\pi$ of the remaining tensor network.

\STATE 2. Construct a variational neural network.

\FOR{$0 \leq i<\mathrm{Epoch}$}

\STATE (a) Draw samples from the nerual network based on Eq.\ref{eq:8}.

\STATE (b) Compute the effective energy $\tilde{E}(\boldsymbol{s})$ of these samples using tensor network contraction.

\STATE (c) Compute the loss function $F_q$ and its gradient based on Eq.\ref{eq:9} and Eq.\ref{eq:10}.
\STATE (d) Using the above gradient update parameters as well as the loss function.
\ENDFOR

Draw samples and compute the variational free energy $F_q$.
\RETURN $F_q$

\end{algorithmic}
\end{algorithm}

\section{Numerical Results}\label{sec:4}
In this section, we present a comprehensive evaluation of TNVAN across diverse spin systems, emphasizing its accuracy, scalability, and adaptability to different topologies. The experiments include both exactly solvable models (e.g., 2D Ising models) and computationally challenging systems (e.g., spin glasses on random graphs and fully connected networks). 

\begin{figure}[htbp]
\centering
    \includegraphics[width=0.5\linewidth]{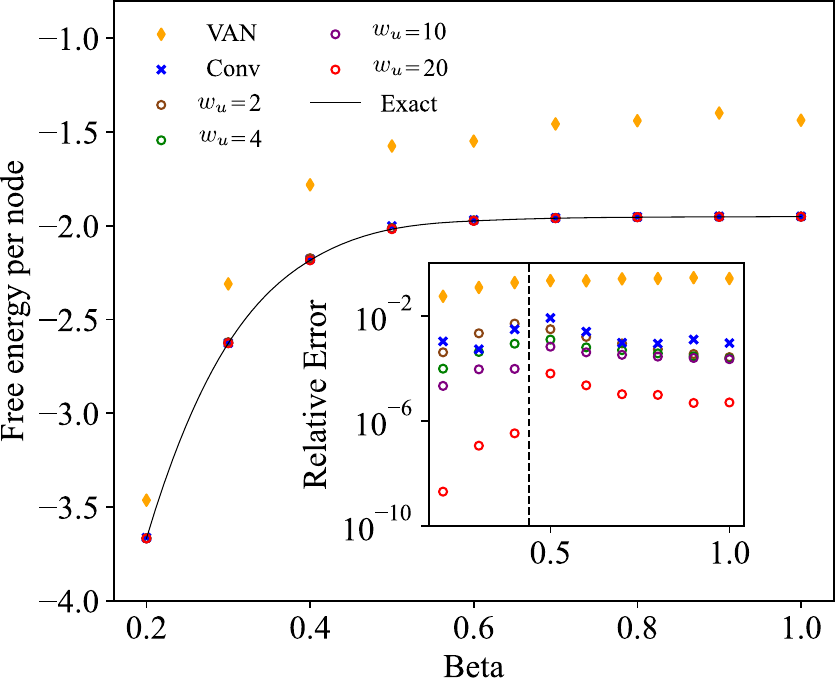}
    \caption{Free energies of $40\times 40$ 2D Ising models. We set $w_u =\{2 , 4 , 10 , 20\}$ and got $|\mathcal{W}| = \{507 , 255 , 106 , 33\} $, representing the number of inputs for the neural networks. The free energies were calculated at various temperatures, and our method was compared with VAN and convolutional VAN. The inset shows the relative errors. The exact results were obtained using the Kac-Ward solution. The hyperparameters of these numerical experiments are provided in Table.~\ref{tab:1}.}
    \label{fig:2}
\end{figure}

We first benchmark TNVAN on the 2D ferromagnetic Ising model with open boundary condition, where the exact free energy can be computed via the Kac-Ward solution~\cite{kac1952combinatorial}. The system size (1,600 spins) poses a significant challenge for traditional tensor network methods due to memory constraints. TNVAN decomposes the lattice into a width set $\mathcal{W}$ and a remaining subsystem $\mathcal{R}$, with the contraction width $w_u$ controlling the trade-off between accuracy and computational complexity. We compared our method with other variational methods within the framework of VAN ~\cite{PhysRevLett.122.080602, PhysRevE.103.012103}, which have demonstrated superior performance compared to other variational methods such as mean field~\cite{jordan1999introduction, thouless1977solution} and belief propagation~\cite{yedidia2003understanding}. For $w_u = \{2, 4, 10, 20\}$, the corresponding width set sizes are 
$|\mathcal{W}|=\{507, 255, 106, 33\}$. As shown in Fig. 2, TNVAN consistently outperforms standard VAN and convolutional VAN (Conv-VAN). (which uses the PixelCNN~\cite{pmlr-v70-kalchbrenner17a} structure to achieve the autoregressive property.) across all temperatures ($T=1.0\ \mathrm{to}\ T=5.0$). At critical temperature
$T_c\approx 2.26918$ where correlations diverge, TNVAN (with $|\mathcal{W}|=33$) achieves a relative error of the order of $10^{-4}$, compared to $10^{-1}$ for VAN and $10^{-2}$ for Conv-VAN. This improvement stems from TNVAN’s ability to project huge 2D lattices onto small width sets, reducing the complexity of variational distribution. Notably, even with $|\mathcal{W}|=507$ TNVAN retains high precision, demonstrating robustness against larger width sets.

A key advantage of TNVAN lies in its parameter efficiency. TNVAN requires fewer parameters than VAN and Conv-VAN (see in Tab.~\ref{tab:1}) to get more accurate results, as the latter introduces convolution layers to capture long-range correlations which is redundant in TNVAN due to tensor networks have already made the system as dense as possible. Furthermore, TNVAN uses the batch-contraction strategy for computing enabling efficient GPU utilization.

\begin{table}[htbp!]
\label{tab:1}
\caption{Hyper parameters of nerual networks.}

\begin{ruledtabular}
\begin{tabular}{ccccc}
\textrm{}&
\textrm{TNVAN}&
\textrm{VAN}&
\textrm{Conv}\\
\colrule
Net Depth & 1 & 1 & 2\\
Channels & 2 & 2 & 1\\
Max Steps & 1000 & 1000 & 1000\\
Batch Size & 1000 & 1000 & 1000\\
Learning Rate & 0.01 & 0.01 & 0.01\\
Number of Inputs & $|\mathcal{W}|$ & 40$\times$40 & 40$\times$40\\
\end{tabular}
\end{ruledtabular}
\end{table}

To evaluate the ability of our approach to estimate the free energy in more complex and sparse systems, we consider the $\pm J$ spin glass models on random graphs.  This model is known as the Viana-Bray spin glass model~\cite{viana1985phase}, which has a distribution
of couplings following $P(J_{i j} = 1) = P(J_{i j} = -1) = 1/2$. We set the inverse temperature $\beta=0.5$ These systems exhibit frustration and complex energy landscapes, making them intractable for exact methods. TNVAN's performance is quantified by the converged variational free energy $F_q$. A lower value of $F_q$ indicates a closer approximation to the true free energy, as guaranteed by the non-negativity of the KL-divergence~\cite{mackay2003information}. 

\begin{figure*}[!htbp] 
	\includegraphics[width=0.9\linewidth]{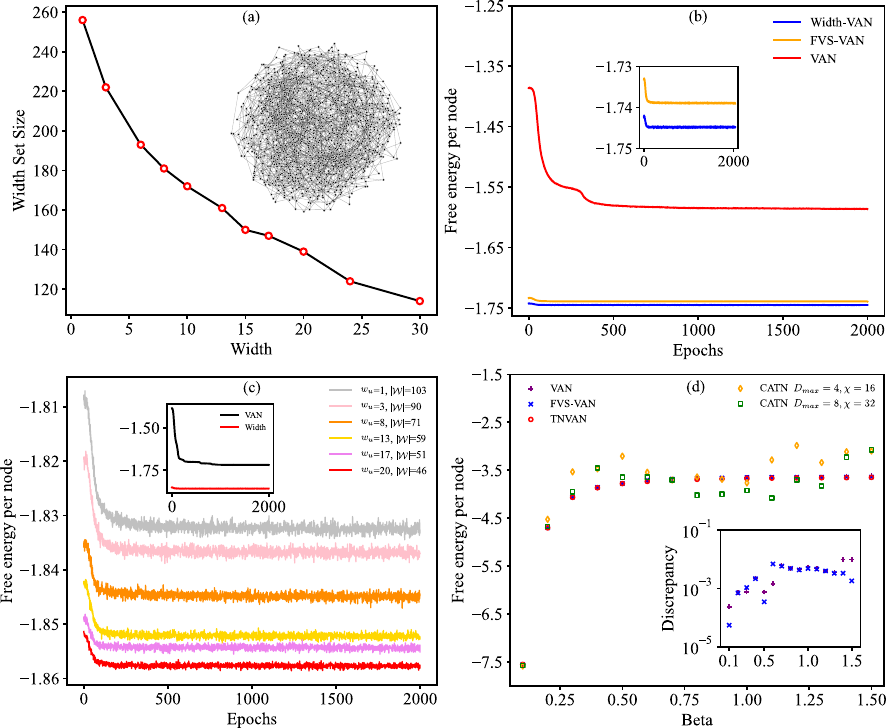}

\caption{Free energies of spin glass systems. (a)  
Size of the width set $|\mathcal{W}|$ for different width upper bounds $w_u$ on a random regular graph with $n=1000$ and $degree = 3$ shown in the inset. When $w_u=1$, the width set is equivalent to the FVS. (b) Changes in the variational free energy with training epochs for the random regular graph in (a). The width upper bound of TNVAN is set to $w_u=20$, leading to a width set size $|\mathcal{W}| = 139$. Compared to VAN and FVS-VAN, TNVAN achieves a lower variational free energy. (c) Impact of the width set size on the TNVAN performance for a random regular graph with $n=1000$ and $degree = 4$. The variational method yields better results as the number of variables decreases. For both models, we set $\beta=0.5$, couplings $J_{ij} = \pm 1$ with distribution $p(J_{ij} = +1)=p(J_{ij} = -1)= \frac{1}{2}$. (d) Sherrington-Kirkpatrick model with $n=50$, and $J_{ij}$ sampled from a standard normal distribution $\mathcal{N}(0,1)$. We calculated the free energy at different temperatures using different variational methods and CATN with different hyper-parameters. Inset is the discrepancy of other variational methods relative to TNVAN. For comparison, the hyper-parameters of all neural networks were identical: depth=1, channels=2, learning rate=0.01, epoch=2000, batch size=2000.}
\label{fig:3}
\end{figure*}

We test our method on the spin glass model defined on random regular graphs with $n=1000$ and $degree=3$. Fig.~\ref{fig:3} (a) shows how the width set size changes with the width limit $w_u$. For $w_u = 1$, the remaining tensor network is a tree tensor network, and the width set contains $256$ variables, which is equivalent to FVS. Increasing $w_u=30$, the corresponding width set size reduces to $|\mathcal{W}| = 114$, which is roughly $10\%$ of the original system size. A lower number of variables directly results in a smaller configuration space and cause a lower variational free energy. As shown in Fig.~\ref{fig:3} (b), we set the $w_u=20$ for the parallel tensor network contractions with a batch equal to 10000 on a single A100 GPU, and accordingly the width set size $|\mathcal{W}|=139$. The significant gap between the variational free energies of TNVAN and FVS-VAN indicates that the relative error of the variational free energy obtained by FVS-VAN with respect to the exact value is not less than $10^{-3}$.

In addition, we investigate the impact of the width set size on the performance of the variational method on $n=300$ and $degree=4$ regular random graphs. In Fig.~\ref{fig:3} (c), we set the upper bound of contraction width $w_u=\{1, 3, 8, 13, 17, 20\}$ and obtained the width set size $|\mathcal{W}| = \{103, 90, 71, 59, 51, 46\}$. For $w_u=1$ the width set has 103 variables and is equivalent to the FVS which is about a third of the original system. This conclusion is consistent with that in FVS-VAN. Even for $w_u=3$ and $|\mathcal{W}|$, TNVAN also has an approximate $3\%$ performance increase compared to FVS-VAN. The rest of the results in this figure further validate our conclusion that the variational free energy decreases with decreasing width set size. From the inset of Fig.~\ref{fig:3} (c), it can be seen that TNVAN has a $15\%$ increase in performance compared to VAN. From the numerical results in above random graph Ising model, we find the size of the width set is about $15\%$ of the original system, at $w_u=20$. Although the size of the witdh set increases with the size of the system, it is still much smaller than the number of variables in the original system and the size of FVS ($30\%$ of the original system). Thus TNVAN has a definite  advantage over VAN and FVS-VAN.

Finally, we apply TNVAN to the Sherrington-Kirkpatrick (SK) model~\cite{sherrington1975solvable} with $n=50$
a fully connected spin glass with couplings $J_{ij}\sim\mathcal{N}(0, 1)$. This system tests the capacity of TNVAN to handle long-range interactions and high-dimensional entanglement. Despite the dense connectivity, TNVAN decomposes the graph into a width set with $|\mathcal{W}| = 20$ for $w_u=20$ and handles the remaining subsystem by tensor network contractions. We calculated the free energy at different temperatures for the SK model with $n=50$ using various variational methods, including TNVAN, VAN, and FVSVAN, as well as Contracting Arbitrary Tensor Networks (CATN)~\cite{PhysRevLett.125.060503} with different hyper-parameters. The results are presented in Fig.~\ref{fig:3} (d), with a detailed comparison of the outcomes obtained by TNVAN and other variational methods provided in the inset. From these results, it can be observed that TNVAN consistently achieves lower variational free energy across all temperatures compared to both VAN and FVS-VAN, demonstrating its superior performance in approximating the true free energy.

Furthermore, the free energy computed by CATN exhibits significant oscillations with temperature, indicating that the tensor network method fails to produce reliable or consistent results for this system. This limitation arises from the inherent challenges tensor networks face in performing effective regularization and constructing global low-rank approximations in high-dimensional and dense systems, such as the SK model. Furthermore, an additional advantage of the variational method is its capacity to estimate the upper bound of the true free energy for systems without exact solutions, implying that a lower variational free energy indicates a closer approximation to the true result. In contrast, tensor networks lack this feature, as they are unable to establish a definitive relationship between their computational outcomes and the actual free energy. These findings underscore the advantages of TNVAN in handling complex systems where traditional tensor network methods struggle to maintain accuracy and stability.

The above results have demonstrated that TNVAN has a significant superiority over other variational methods in estimating free energy. Moreover, TNVAN can also be utilized to calculate other physical quantities, such as correlation functions, as introduced in FVS-VAN~\cite{PhysRevE.103.012103}. We will perform these calculations in future work.

\section{Conclusion}\label{sec:5}
 
In this study, we introduce the TNVAN approach, a novel framework that integrates the tensor networks and autoregressive neural networks for computing the free energy of statistical models. TNVAN addresses the limitations of traditional methods by decomposing the system into two distinct parts: the width set and the remaining variables. The width set, a technically selected subset of indices (e.g., spins), plays a critical role in controlling the contraction complexity of the tensor network. By imposing a width bound and identifying the corresponding width set, TNVAN efficiently processes the remaining sparse subsystem, projecting its contributions onto the width set through tensor network contractions. For systems with small width sets, one can employ an exhaustive enumeration approach to obtain exact results, as demonstrated in prior work~\cite{Gray2021hyperoptimized}. For larger width sets, where exact enumeration becomes computationally infeasible, TNVAN leverages a variational autoregressive network to approximate the free energy. The VAN parameterizes the variational distribution of the dense subsystem, combining the expressive power of neural networks with efficient sampling, as shown in~\cite{PhysRevLett.122.080602}.

Our extensive numerical experiments highlight TNVAN's superior accuracy in free energy estimation compared to existing methods, including various neural network-based variational approaches. TNVAN demonstrates robust performance across a wide range of system topologies, such as low-dimensional lattices, random graphs, and the full-connection system with long-range interactions, where traditional methods often struggle. A key advantage of TNVAN is its ability to avoid the computationally expensive singular value decomposition (SVD), which is a common bottleneck in tensor network methods. Instead, TNVAN leverages GPU acceleration to perform efficient batch contractions and gradient-based optimization, significantly enhancing computational performance. This combination of accuracy, scalability, and efficiency underscores TNVAN's potential to address complex statistical mechanics problems that are intractable for conventional approaches.

In future work, we will focus on extending TNVAN to quantum systems, where the entanglement and high-dimensional state spaces cause challenges. Additionally, we aim to optimize TNVAN's adaptive partitioning scheme to dynamically adjust the width set based on system-specific properties, further improving its accuracy and efficiency. By continuing to refine and expand TNVAN's capabilities, we envision it becoming a versatile tool for tackling a broad of problems in statistical mechanics, quantum physics, and beyond.
\begin{acknowledgments}
This work is supported by Projects 12325501, 12047503, and 12247104 of the National Natural Science Foundation of China and Project ZDRW-XX-2022-3-02 of the Chinese Academy of Sciences. P.~Z. is partially supported by the
Innovation Program for Quantum Science and Technology
project 2021ZD0301900.
\end{acknowledgments}


\bibliography{main.bib}

\appendix

\end{document}